\documentclass[sigconf,nonacm=true]{acmart}

\setcopyright{rightsretained}
\copyrightyear{2020}

\usepackage{natbib}
\usepackage{wrapfig,lipsum,booktabs} 
\usepackage{sidecap}    
\usepackage{multirow}	
\usepackage{textcomp}	
\usepackage{pifont}		
\usepackage{graphicx}

\usepackage{pgfplots}	
\pgfplotsset{compat=1.8}
\usepgfplotslibrary{statistics}

\usepackage{listings} 
\usepackage{color}
\definecolor{darkgreen}{rgb}{0,0.5,0}
\definecolor{darkred}{rgb}{0.5,0,0}
\lstdefinelanguage{diff}{
  basicstyle=\ttfamily\scriptsize,
  morecomment=[f][\color{blue}]{@@},     
  morecomment=[f][\color{darkgreen}]+,       
  morecomment=[f][\color{darkred}]-,         
  morecomment=[f][\color{grey}]{---},    
  morecomment=[f][\color{grey}]{+++},    
}

\newcommand*{\ignore}[1]{}

\newcommand*{\register}[1]{\texttt{\%#1}}
\newcommand*{\registerOne}[2]{\register{#1}\texttt{[#2]}}
\newcommand*{\registerTwo}[3]{\register{#1}\texttt{[#2:#3]}}
\newcommand*{\simplex}{\textsc{Simplex}}
\newcommand*{\simplexurl}{\url{https://github.com/bingseclab/simplex}{}}

\begin{document}

\title[Simplex: Repurposing Intel MPX]{\simplex{}: Repurposing Intel\textregistered~Memory Protection Extensions for Information Hiding}
\subtitle{Salvaging Endangered Hardware Features for Security}
\author{Matthew Cole}
\email{mcole8@binghamton.edu}
\orcid{0000-0003-1743-1504}
\affiliation{%
	\institution{Binghamton University}
	\city{Binghamton}
	\state{New York}
}

\author{Aravind Prakash}
\email{aprakash@binghamton.edu}
\affiliation{%
	\institution{Binghamton University}
	\city{Binghamton}
	\state{New York}
}

\begin{abstract}
	With the rapid increase in software exploits, the last few decades have seen several hardware-level features to enhance security (e.g., Intel MPX, ARM TrustZone, Intel SGX, Intel CET). 
	Due to security, performance and/or usability issues these features have attracted steady criticism. 
	One such feature is the Intel\textregistered~Memory Protection Extensions (MPX), an instruction set architecture extension promising spatial memory safety at a lower performance cost due to hardware-accelerated bounds checking.
	However, recent investigations into MPX have found that is neither as performant, accurate, nor precise as cutting-edge software-based spatial memory safety.
	As a direct consequence, compiler and operating system support for MPX is dying, and Intel has begun to manufacture desktop CPUs without MPX. 
	Nonetheless, given how ubiquitous MPX is, it provides an excellent yet under-utilized hardware resource that can be aptly salvaged for security purposes. 
	In this paper, we propose \simplex{}, a library framework that re-purposes MPX registers as general purpose registers.
	Using \simplex{}, we demonstrate how MPX registers can be used to store sensitive information (e.g., encryption keys) directly on the hardware. 	
	We evaluate \simplex{} for performance and find that its overhead is small enough to permit its deployment in all but the most performance-intensive code.
	We refactored the \texttt{string.h} buffer manipulation functions and found a geometric mean 0.9\% performance overhead.
	We also modified the \texttt{deepsjeng} and \texttt{lbm} SPEC CPU2017 benchmarks to use \simplex{} and found a 1\% and 0.98\% performance overhead respectively.
	Finally, we investigate the behavior of the MPX context with regards to multi-process and multi-thread programs. 
	
\end{abstract}

\maketitle


\section{Introduction}
\label{sec:introduction}

Intel\textregistered~Memory Protection Extensions (MPX) is an instruction set architecture (ISA) extension for modern Intel processors providing spatial memory safety using compile-time intentions.
MPX is comprised of three key components working in harmony: architectural support through a set of two configuration, one status, and four bounds registers; compile-time instrumentation; and run-time support integrated with the operating system.
This run-time manages enabling and disabling CPU interpretation of MPX instructions through the configuration registers, sets up the pointer bounds lookup table, interprets error codes indicated in the status register, and coordinates with the operating system to handle memory management and error handling. 

In practice, MPX is unusable in its intended form. It was intended to be performant, interoperable with uninstrumented legacy code, and configurable for both debug and release environments without rewriting the source.
However, Oleksenko et al. and Serebryany independently showed that MPX does not perform as well as software- and language-based memory safety, demonstrating a 50\% amortized performance overhead with good compiler optimizations, and a 400\% worst-case performance overhead \cite{Oleksenko2017,Serebryany2016}.

\ignore{Worse yet, even though Intel claims that the MPX instruction subset is interpreted as a NOP on legacy processors, Oleksenko \etal{} still found a 15+\% slowdown.
At this time, only two compilers \--- Intel C Compiler (ICC) and the Visual Studio 2015 Compiler \--- support compiling programs with MPX instrumentation and with MPX-wrapped standard library functions.}
As such, MPX has steadily lost support in the software community.
The GNU C Compiler (GCC) recently removed its libmpx library and eliminated the instrumentation code, while Clang has never supported MPX beyond specifying the instruction opcodes in its x86 targets.
Additionally, 
Linux recently removed its support for kernel compilation with MPX because of the lack of support from the GNU community.
In short, it now appears that MPX will not achieve widespread adoption as a memory safety tool that it was designed for. 
Yet, MPX is already a supported feature on widely deployed processors (e.g., Skylake). 
For example, Passmark reported that by 2017 all of the Intel CPUs that they sold were the first-generation Skylake model or later~\cite{Passmark2019}.
Even a conservative estimate puts the number of MPX-supported deployments at 100s of millions worldwide.
Therefore, MPX is a ubiquitous---yet unused---resource. 

In this paper, we leverage MPX for general purpose data storage with emphasis on data hiding. 
Ability to hide data is a valuable resource in security. 
For example, in the enforcement of control-flow integrity using a shadow stack, it is necessary to ensure integrity of the shadow stack while accommodating frequent updates to the shadow stack whenever there is a {\tt call} or a {\tt ret} instruction. 
Similarly, it is often beneficial to isolate critical encryption keys and passwords from memory in order to protect their integrity and confidentiality in case of memory corruption and/or information leakage.
On the one hand, hiding data in the kernel is often impractical as it incurs (sometimes prohibitive) performance overhead due to the expensive transition to/from user and kernel modes. 
But on the other hand, recent attacks demonstrate that hiding data in userland is ineffective even in a 64-bit address space\cite{Oikonomopoulos2016}. 

Our contribution is codenamed \simplex{},
which is comprised of a library enabling extrospection and manipulation of the MPX context, a minimalist runtime that avoids the overhead associated with the compiler-provided MPX runtime, a test suite verifying correctness, and evaluations demonstrating the practicality of \simplex{}.
\simplex{} provides abstractions to the underlying MPX operations such that its ``bounds" registers appear to behave identically to general purpose registers, even though Intel does not provide specific access and mutation instructions in the MPX extension.
\simplex{} completely avoids the bounds lookup table and associated runtime costs, which form the main source of overhead for MPX use~\cite{Oleksenko2017,Otterstad2015}. 
\simplex{} provides storage in MPX registers optimally used to store data that is undesirable or risky to be stored in userspace memory, when the data size is small (ideally 8$\times$64 bits, although \simplex{} also provides 4$\times$128-bit operations), and when the data is infrequently accessed.
\simplex{} is particularly beneficial to use in cases such as information hiding because previous works use a modified compiler to reserve a general purpose register for hiding (e.g.~\cite{Kuznetsov2014,Lu2015,Mohan2015,Zhang2015}).
Reserving registers is undesirable for two reasons: (1) it removes a register from the allocation pool, which could in-turn impact performance due to sub-optimal register allocation~\cite{Bruening2000}, and (2) it affects interoperability when handwritten assembly or binaries not compiled using the modified compiler may accidentally access or modify the reserved register.
This in turn may compromise confidentiality or integrity of the hidden data.
Because \simplex{} uses the MPX bounds registers, and because the bounds registers are not used unless the application was also explicitly compiled with MPX support, we can ensure that no other code will access or modify the hidden data or pointer stored inside the bounds register.


\ignore{We believe that MPX is ubiquitous based on inferences from publicly available data. 
Based on publicly-available data from a variety of sources in both the server and desktop segments, we find that between about three in four and nine in ten computers has an Intel processor.
We also found that retailers overwhelmingly sell only the most recent version of both Intel and AMD processors.
One retailer reported that by 2017 all of the Intel CPUs that they sold were the first-generation Skylake model or later, and that by 2019, only 1\% of their sales were from that first-generation Skylake model.
Given that most experts agree that the average desktop computer lasts between three to five years and best estimates of the number of PCs worldwide is over 1.3 billion.
We therefore conservatively estimate that by making \simplex{} available, hundreds of millions of computers would gain the ability to run software taking advantage of specialized storage suitable for a variety of security-sensitive applications including information hiding or cryptographic operations.
}

Our evaluation shows that \simplex{} is practical, and confirms initial observations by Otterstad~\cite{Otterstad2015} and Oleksenko et al.~\cite{Oleksenko2017} that the majority of MPX's performance cost comes from interacting with the bounds lookup table and associated costs within the runtime.
We avoid this overhead because \simplex{} avoids using the bounds lookup table by writing to the bounds registers directly using the \texttt{bndmk} instruction and reading from the bounds registers using the \texttt{bndmov} instruction to spill the contents into memory.
We evaluated for performance in two different ways.
First, we created three custom benchmark fixtures: a microbenchmark testing the rate at which load and store operations can be completed to both the \register{r15} general purpose register and the \register{bnd0} MPX register, a macrobenchmark simulating information unhiding by traversing and combining two hidden half-buffers, and a series of benchmark implementations of memory operations from the \texttt{string.h} header.
Second, we compiled sandboxed versions of two SPEC CPU2017 benchmarks: 519.lbm, a particle-fluid simulation written in C, and 531.deepsjeng, a chess engine written in C++.
Finally, we evaluated for usability and correctness by modifying the OpenSSL Blowfish cipher, then running the included integration and unit test suites.
We further discuss these evaluations in \S\ref{sec:evaluation}.

\ignore{
The remainder of the paper is structured as follows:
We discuss the history of MPX and the reasons prohibiting its widespread adoption as a memory safety tool in \S\ref{sec:background:history mpx}. 
We next examine the problems in information hiding continuing to plague security researchers in \S\ref{sec:background:information hiding}.
We provide an overview of our threat model and necessary modifications to a compiler to support \simplex{} in \S\ref{sec:overview}.
We describe the implementation of the \simplex{} library, and answer questions about MPX context behavior during common program behavior including multithreading and system process lifecycles \S\ref{sec:implementation}.
We present our evaluation in \S\ref{sec:evaluation}, showing \simplex{} is both sound and practical.
We survey related work in \S\ref{sec:related work} and conclude in \S\ref{sec:conclusion}.
}

\section{Background}
\label{sec:background}

\subsection{Intel MPX}
\label{sec:background:history mpx}
In 2012, Intel introduced \textsc{PointerChecker}, which provides bounds checking in the software layer through the Intel Composer XE development environment for C and C++~\cite{Ganesh2012}.
Recognizing the potential for greatly improved performance through hardware support, Intel moved much of the \textsc{Pointer Checker} functionality into MPX, announced in 2013~\cite{Intel2013} and subsequently debuted in the Skylake architecture in 2015. 

MPX is a combination of an instruction set extension, compiler and operating system support, and runtime library.
It provides four new 128-bit bounds registers (\register{BND0} through \register{BND3}), each of which are split into an upper half and lower half which have the purpose of holding an upper bound and lower bound address.
MPX also employs the \register{BNDCFGx} register pair to hold user-space and kernel-space configuration, and a \register{BNDSTATUS} register to hold status information in case of a bounds check failure.
Intel designed MPX with the overarching goal of compatibility with uninstrumented code and unextended architectures. 
Where an MPX-supported CPU encounters uninstrumented code, such as a vendor-provided library, program execution continues with the cost that the CPU can no longer provide memory safety because the bounds checks are not performed unless a \texttt{bndcl}, \texttt{bndcu}, or \texttt{bndcn} instruction is executed.
Where an MPX-unsupported CPU encounters instrumented code, or when MPX has not been initialized by setting \registerOne{bndcfg}{0}, the instructions are interpreted as \texttt{nop} instructions instead of triggering unrecognized instruction exceptions.

\begin{figure*}[!hbt]
	\includegraphics[width=0.75\textwidth]{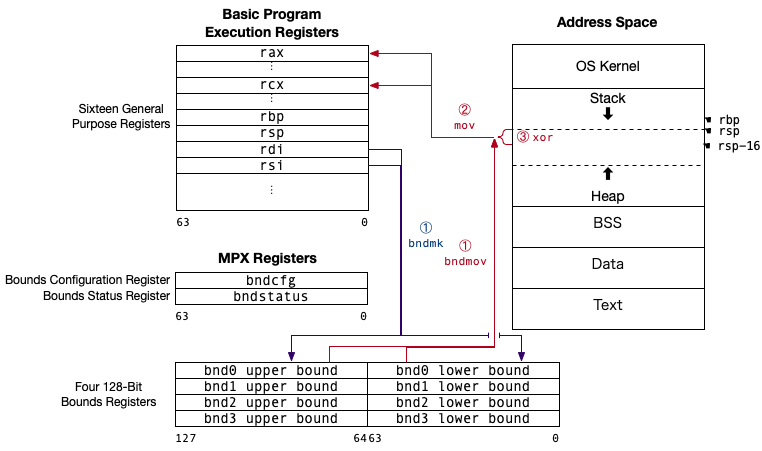}
  \caption{The MPX context as part of the larger Intel64 context. The blue pathway shows how information is written to the bounds registers. The red pathway shows how information is read from the bounds registers (including sanitizing the stack afterwards).}
  \label{fig:mpx-context}
\end{figure*}

\subsection{Ubiquity of MPX}
\label{sec:background:ubiquity}
Although no direct count of deployed processors with MPX support exists, we nonetheless believe that MPX is ubiquitous based on inferences from publicly available data.
For example, the Steam Hardware and Survey for May 2019 \--- which automatically polls the hardware of the Steam gaming service's users \--- shows that 81.97\% of users have an Intel processor~\cite{Steam2019}.
Likewise, PassMark reported that based on the benchmark baselines submitted from July 2013 (when MPX was introduced) through the present, Intel has had a market share between 73.60\% and 82.50\%~\cite{Passmark2019}.
This is an important contrast to the Steam survey because it does a better job of accounting for the server market, whereas the Steam survey highlights the desktop market, specifically gaming computers.
More optimistically, Intel's 2018 financial reports indicated that they held 84.2\% of the desktop market, 87.9\% of the notebook market, and 96.8\% of the server market, for a weighted share of 90.41\%~\cite{Jeyaratnam2019}.

Second, within Intel's large market share, we believe that the overwhelming majority are compatible with MPX based on sales data.
For example, according to Mindfactory \--- a leading German computer parts retailer \--- not later than 2017, all of the Intel processors that they sold were of the first MPX generation \textit{Skylake} architecture or more recent.
By 2019, \textit{Skylake} represented only 1\% of Mindfactory's sales, with 92\% belonging to the most recent \textit{Coffee Lake} or \textit{Coffee Lake}-Refresh architectures~\cite{Ingebor2019}.
This suggests that the CPU sales dynamic heavily favors recent models, and thus ISA extensions are quick to penetrate the market share.
Lastly, we note a 2016 estimate that by 2019 there would be over 1.3 billion desktop computers worldwide~\cite{Forni2019}.
Thus it is reasonable to conclude that the potential impact of \simplex{} is high as numerous computers \--- possibly hundreds of millions \--- will gain register storage through repurposing their bounds registers.
This benefit comes without modification to the underlying architecture or perceptible cost to the end user and has security-sensitive applications including information hiding or cryptographic operations.

\subsection{Present State of MPX}
\label{sec:background:present state}
MPX is impractical to use in its intended form. Although MPX achieves a four- to five-fold speedup compared to \textsc{Pointer Checker}~\cite{Oleksenko2017}, it is not without significant issues hindering its widespread adoption, foremost of which is the execution cost.
MPX-enabled benchmarks experience worst-case 200\% performance overhead, 480\% memory overhead and $5.4x$ more page faults~\cite{Oleksenko2017}.
Additionally, bounds table lookups appear to cause significant cache pressure.
Even on legacy hardware, where the MPX instructions are interpreted in an idempotent manner, there is still up to a 50\% slowdown~\cite{Serebryany2016}.
Furthermore, MPX \ignore{is not fully usable and interoperable as a memory safety defense.
It }cannot catch temporal memory safety issues such as heap use-after-free and stack use-after-return~\cite{Oleksenko2017}, it has false positives from otherwise legal C idioms due to restrictions on structure memory layouts~\cite{Oleksenko2017,Serebryany2016}, it experiences false negatives in response to undefined behaviors which cause inappropriate bounds loads~\cite{Otterstad2015}, it conflicts with other Intel ISA extensions such as SGX and TSX~\cite{Oleksenko2017}, and it has no explicit support for multithreading~\cite{Oleksenko2017}.

As a result, support for MPX has waned\ignore{even as support for Address Sanitizer~\cite{Serebryany2012} has grown, which has been officially supported since GCC 4.8, Clang 3.1, XCode 7.0, Visual Studio 2019 16.4, and Linux kernel 4.0}.
Currently MPX's only compiler support is Intel's own ICC since version 15.0 and Microsoft's Visual Studio 2015 Update 1.
GCC has dropped support and Clang never supported MPX. 
\ignore{
Although GCC introduced support in GCC 5.0, this was removed in GCC 9.0 and Clang has never supported MPX bounds checking although it does support code generation of the MPX instructions. 
Runtime support for MPX is likewise waning: only Microsoft Windows 10 provides a daemon runtime component after Linux kernel 5.6 and QEMU hypervisor 4.0 removed support for MPX.
Therefore we propose re-purposing MPX to address other issues facing the security community rather than leave the bounds registers unused or to deprecate them by a future microcode update.
}

Despite the sunset of support for MPX in its memory safety usage, we must emphasize that \simplex{} does not rely on either compiler or operating system support to function.
The \simplex{} library provides all necessary runtime components and functions for instrumentation, and the MPX context is part of the broader XSAVE context, thus it is still saved and restored on context switches even though Linux formally removed all MPX support as of kernel version 5.6.

\subsection{Information Hiding}
\label{sec:background:information hiding}

Recent works demonstrate that information hiding techniques relying on probabilistic mechanisms can be defeated.
G\"{o}kta\c{s} et al. demonstrated \textit{thread spraying}~\cite{Goktas2016} as a means of disclosing the safe regions with a known structure.
By repeatedly creating objects that have safe stacks and regular stacks, then probing the space to find one of these hidden safe stacks, they can effectively de-randomize the address space.
They also discovered that information in the thread local storage (TLS) and the thread control block (TCB) provide clues to locating these stacks. 
Furthermore, Oikonomopoulos et al. introduced \textit{allocation oracles} which eliminate the need for probing~\cite{Oikonomopoulos2016}.
The idea is that an allocation oracle takes the size of an area to allocate as input, and if successful returns the location allocated.
From this information and applying a binary search technique, an attacker can locate ``holes'' in the allocatable memory.
If the attacker has knowledge of how a defense's sensitive data is laid out, then these holes reveal where the sensitive data \textit{is not} hidden.
With enough queries to the oracle, eventually the sensitive data can be located, and the process avoids crashes or distinguishable behavior usable by a runtime detector.
Likewise, Evans et al. used timing side channels to read the contents of hidden metadata with or without crashes (the former is faster, the latter is difficult to detect)~\cite{Evans2015}.
Using this technique, they can de-randomize the location of libraries such as libc, then use this to calculate the start of the safe region.
Once complete, modifying the contents of the safe region permits an attacker to violate at least one implementation of CPI.

Recent work has shown that having registers to simulate segmentation as available in the IA-32 architecture can be used to provide deterministic rather than probabilistic information hiding.
Koning et al. introduce \textsc{MemSentry}, a collection of implementations of information hiding relying on a variety of hardware support~\cite{Koning2017}.
One common point of these implementations is that they would benefit from dedicated registers.
Finally, two of the implementations of Code Pointer Integrity require a dedicated register for information hiding~\cite{Kuznetsov2014}.
In the implementation released at the time of publication, the \register{fs} register was reserved, however this may affect other legitimate usages of the register.
For example, operating systems sometimes use this register to access thread local storage (TLS).
Providing register storage via \simplex{}  helps return reserved general purpose registers to the compiler's allocation list and restores special purpose registers to their expected usage.


\section{\simplex}
\label{sec:overview}

\subsection{Threat Model}
We assume a threat model similar to that offered by other work on information hiding, namely Koning~\cite{Koning2017} and Yun~\cite{Yun2019}.
Our system under threat has an effective defense against code reuse, which in turn prevents an attacker from arbitrarily calling the \simplex{} library functions, even though he or she may have an arbitrary read or write primitive.
Although \simplex{} might be used to store a pointer to a \textit{hidden} memory region, it does not itself provide \textit{isolation}.
We presume that the programmer has a \textit{Trusted Code Base} comprised of at least a privileged, trusted operating system and a trusted build toolchain used to build the \simplex{} library.
We concede that an attacker may be able to load a Loadable Kernel Module (LKM) that enables or disables MPX at a privileged operating system level (and in fact, we provide one such implementation within the \simplex{} code base).
However, this would imply a compromised kernel, which is outside our scope. 
That said, we show in \S\ref{sec:context behavior}, that is not sufficient for an attacker to emplace values into the bounds registers or leak values from the bounds registers in a way that is beneficial to the attacker.
Finally, we assume that \simplex{} is correctly implemented and is trusted by the programmer.
We release our code as open source, and offer a full test suite within that code base as an assurance to that assumption.

\subsection{Design Decisions}
Previous works seeking to hide information from attackers have chosen one of three options.
1) Storing information in the kernel or in pages that can only be accessed in a privileged hardware mode (e.g. \cite{Intel2019,Gruss2017}) is secure as long as the operating system is not compromised.
However these schemes come with the obligation of additional context switches for each query or update, hampering performance.
2) A more performant choice is storing information in a hidden region within the program's address space (e.g. \cite{Davi2015,Kuznetsov2014,Mohan2015}).
Yet it relies on either probabilistic hiding measures which can be defeated if the attacker has knowledge of the type of information being hidden, or if the attacker is able to tolerate crashes and restarts while searching.
3) Alternatively, it is possible to reserve registers from the compiler's allocation pool and use these registers exclusively for storing sensitive data.
Once the registers are selected, the defender can formally verify that no other code accesses these registers, guaranteeing security.
Nonetheless, there is still the concern that available registers are limited and may conflict with other defenses or dynamically linked code that use the reserved register.

\subsection{Simplex-Enabled Compilation}
In our evaluations, we manually replaced global pointer objects and their reference/dereference statements with the necessary code to enable bounds register usage.
However, we do not feel this is scalable.
Consider the modifications made to the SPEC CPU2017 benchmarks:
\texttt{519.lbm} has just 1 KLOC and required 22 modifications, \texttt{531.deepsjeng} has only 10 KLOC and required 173 modifications \-- these are very small code bases compared to \texttt{502.gcc} (1.3 MLOC) and \texttt{526.blender} (1.6 MLOC), the largest C/C++ benchmarks in CPU2017.
Making these modifications are expensive in terms of developer effort and time, requiring both discovering and understanding the global variables' utilization.
For example, modifying the two SPEC CPU2017 benchmarks took about two days of development time each.
If the number and complexity of changes necessary were to scale, implementing the larger benchmarks by hand could take months!
Therefore, we have designed but not yet implemented a system using Clang's annotation system to mark variables as candidates for placement in a bounds register.
This reduces the developer's workload to simply recognizing which variables should go into a bounds register, applying annotations to the declarations, then compiling the source code with the options necessary to enable \simplex{}.

First, the developer applies the necessary annotation at the variable's declaration.
The compiler recognizes the annotation, and maps that variable to one of the bounds registers, depending on its size or throws a compilation error if no more register space is available.
Next, the compiler pass replaces references to these variables with appropriate \simplex{} function calls.
If the variable is an lvalue, it is replaced with a call to one of the mutator functions; if it is a rvalue, it is replaced with a call to one of the accessor functions.

\vspace{.08in}
\noindent
{\bf Developer annotation vs Automated discovery:} On the one hand, developer annotation has the benefit of precisely capturing what is of security relevance and importance as per software design, but on the other hand, developers are prone to make mistakes. Therefore, we recommend 3 modes of operation that makes a trade off between security and performance.  

\vspace{.04in}
\noindent
{\em Whitelisting:} In this mode, we allow a developer to whitelist security-sensitive data that is stored in the MPX bounds registers by the compiler. This is the most conservative and performance-friendly, yet error-prone option.

\vspace{.04in}
\noindent
{\em Automatic inference:} In this mode, the compiler employs a heuristic approach to automatically profile and identify security sensitive information and accordingly provisions MPX bounds registers to manage such sensitive data. One option is to identify security-sensitive documented API functions and perform backward slicing to identify data of interest. This is the most aggressive option that favors security over performance.  

\vspace{.04in}
\noindent
{\em Blacklisting:} Finally, as an intermediate option, blacklisting allows a developer to define data items that {\em should not} be stored in the MPX registers. While blacklisting is just as prone to human error as is whitelisting, it is likely to have less adverse effects on security as compared to mistakes in whitelisting.

\subsection{Context Behavior}
\label{sec:context behavior}

We performed experiments to determine the behavior of the MPX context using the system under evaluation described in \S\ref{sec:evaluation}.
We identified three phases in the lifespan of a scheduling entity: its creation (i.e. calling \texttt{fork()} for processes, or \texttt{pthread\_create()} for threads), mid-life \--- specifically when a parent or sibling modifies its own context \--- and deletion (i.e calling \texttt{exit()} and \texttt{wait()} for processes, or \texttt{pthread\_wait()} and \texttt{pthread\_join()} for threads).

\subsubsection{Processes}
According to the POSIX standard, upon calling \texttt{fork()}, a new process is created by duplicating the parent.
As part of this duplication, the child has a memory space, processor context and file descriptor table that are initially identical but separate from the parent.
However, the child has a unique process ID (and thus is scheduled independently from its parent). Furthermore, the child does not inherit the parent's memory locks, signals, semaphores, processor timers or counters.

At process creation, the child inherits an identical MPX context to that of the parent because the MPX context is itself part of the larger CPU context (see Figure \ref{fig:mpx-context}).
If MPX is enabled or disabled because of the values on the \registerTwo{BNDCFGx}{1}{0} bits for the parent, it will be likewise enabled or disabled for the child upon its creation via \texttt{fork()}.
Additionally, the values in the parent's bounds registers will be inherited by the child's bounds registers, because the bounds registers are a component of the MPX context.
Meanwhile, during the lifespan of both the child and the parent, changes in one process' context do not affect another process even if one is an ancestor or sibling of the other.
Subsequently disabling or enabling MPX in a parent does not confer this change in status to the child, nor does changing the values of a parent's bounds registers propagate to the child.
Furthermore, terminating one process does not change the MPX context of any related process.

The reader may question whether a child might inadvertently pass information about its \simplex{} state to the parent through \texttt{wait()}.
At a minimum, \texttt{wait()} passes the process id through its return value. 
Optionally, it may pass a developer specified integer values using \texttt{wstatus} from child to parent, and may pass an integer error code through the \texttt{errno} flag.
However no processor context is contained in this information and therefore two processes using \simplex{} are fully independent from each other beginning immediately after process creation.
Thus there is no risk of leaking secrets across processes when using \simplex{} above and beyond the risk that is incurred by hiding that information in the process address space or in a reserved register.

Briefly, upon creation of a process, the child {\it inherits} the context of the parent, but this context is {\it unshared} thereafter.
We summarize these findings in Table~\ref{tab:processes}.
\begin{table*}[!htb]
\centering
\begin{tabular}{lllll}
\hline
 & \multicolumn{2}{c}{Parent} & \multicolumn{2}{c}{Child} \\ \hline
Event & Enabled? & BND0 & Enabled? & BND0 \\ \hline
\begin{tabular}[c]{@{}l@{}}Parent calls \texttt{process\_specific\_init()}\\ and \texttt{setbnd(BND0,1)}\end{tabular} & \ding{51} & 1 &  &  \\
Parent calls \texttt{fork()} & \ding{51} & 1 & \ding{51} & 1 \\
Child calls \texttt{setbnd(BND0,2)} & \ding{51} & 1 & \ding{51} & 2 \\
Child calls \texttt{process\_specific\_finish()} & \ding{51} & 1 & \ding{55} & - \\
Child calls \texttt{exit()} & \ding{51} & 1 &  &  \\
Parent calls \texttt{process\_specific\_finish()} & \ding{55} & - &  &  \\ \hline
\end{tabular}
\caption{\simplex{} context behavior for a parent and child process.}
\label{tab:processes}
\end{table*}

\subsubsection{Threads}
POSIX threads are a kernel scheduling entity, such that a single process contains multiple threads, all of which are executing the same program.
Thus although each process' threads share global memory, the threads must have their own call stack and its own CPU context in order to maintain its own program counter.
When a process creates a new thread using \texttt{pthread\_create()}, the newly created thread initially inherits the calling process' CPU state, except that in order to begin execution at the specified start routine, the program counter must be set to the appropriate address.
Accordingly, the newly created thread inherits the calling process' MPX configuration and bounds registers context, similar to the inheritance between a parent and child process described above.
Similar to the behavior shown by processes, the various threads of a single process do not share the configuration nor bounds registers' values once the thread is initialized; the created thread's MPX context is immediately independent from that of its creator.

Briefly, upon creation of a thread, the child {\it inherits} the context of the parent, but this context is {\it unshared} thereafter.
We summarize these findings in Table~\ref{tab:threads}.
\begin{table*}[]
\centering
\begin{tabular}{@{}lllllll@{}}
\toprule
 & \multicolumn{2}{c}{Parent} & \multicolumn{2}{c}{Child 1} & \multicolumn{2}{c}{Child 2} \\ \midrule
Event & Enabled? & BND0 & Enabled? & BND0 & Enabled? & BND0 \\ \midrule
\begin{tabular}[c]{@{}l@{}}Parent calls \texttt{process\_specific\_init()}\\ and \texttt{setbnd(0,0)}\end{tabular} & \ding{51} & 0 &  &  &  &  \\
\begin{tabular}[c]{@{}l@{}}Parent calls \texttt{pthread\_create()}\\ for Child 1\end{tabular} & \ding{51} & 0 & \ding{51} & 1 &  &  \\
\begin{tabular}[c]{@{}l@{}}Parent calls \texttt{pthread\_create()}\\ for Child 2\end{tabular} & \ding{51} & 0 & \ding{51} & 1 & \ding{51} & 2 \\
Child 1 calls \texttt{setbnd(0,1)} & \ding{51} & 0 & \ding{51} & 1 & \ding{51} & 2 \\
Child 2 calls \texttt{setbnd(0,2)} & \ding{51} & 0 & \ding{51} & 1 & \ding{51} & 2 \\
Child 2 calls \texttt{process\_specific\_finish()} & \ding{51} & 0 & \ding{51} & 1 & \ding{55} & - \\
Child 1 calls \texttt{process\_specific\_finish()} & \ding{51} & 0 & \ding{55} & - & \ding{55} & - \\
\begin{tabular}[c]{@{}l@{}}Child 1 and Child 2 each call \\ \texttt{pthread\_exit()}, Parent calls \\ \texttt{pthread\_join()} for each child\end{tabular} & \ding{51} & 0 &  &  &  &  \\
Parent calls \texttt{process\_specific\_finish()} & \ding{55} & - &  &  &  &  \\ \bottomrule
\end{tabular}
\caption{\simplex{} context behavior for a parent and two child threads.}
\label{tab:threads}
\end{table*}

\subsubsection{Repetitive Initialization and Finalization}
Because \simplex{} provides methods to initialize and finalize its minimal MPX context, the reader may question what would happen if a programmer or attacker called these methods repeatedly (whether by accident or malice).
We created a toy program that tests these corner cases.

First, we initialize the program, setting the bounds registers to known values, then initialize the program a second time.
We found that each time the MPX context is initialized, the bounds registers' lower bounds are set to the system maximum unsigned value, and the upper bounds are set to 0.
In MPX's design use case, this results in a guaranteed passed bounds check until the bounds register is set to some allocated object's bounds.
In the \simplex{} use case, repeated initialization destroys the values inside the bounds registers by resetting them to the conservative bounds values.
Although this may allow an attack against availability, it does not allow an attack seeking disclosure.
Furthermore, it is no more dangerous than the numerous \texttt{xor reg reg} gadgets which are used by the compiler to place a zero value in a register.

We also found that after finalization, each of BND1 through BND3 are reset.
However, BND0 displays unusual behavior in that while the lower bound is reset, the upper bound receives a random large value (the most significant bit is always set, but otherwise there is no identifiable behavior).
Thus subsequent repeated finalizations are similar to repeated initializations in that they can cause availability but not disclosure problems.
We summarize these findings in Figure~\ref{tab:repeat}.
\begin{table*}[!htb]
\begin{tabular}{c|c|c|c|c|}
\cline{2-5}
\multicolumn{1}{l|}{} & \multicolumn{2}{c|}{Initialization} & \multicolumn{2}{c|}{Finalization} \\ \cline{2-5} 
\textbf{} & \textbf{First} & \textbf{Subsequent} & \textbf{First} & \textbf{Subsequent} \\ \hline
\multicolumn{1}{|c|}{Config Registers} & Enabled & Enabled & Disabled & Disabled \\ \hline
\multicolumn{1}{|c|}{Bounds Registers} & Reset & Reset & \begin{tabular}[c]{@{}c@{}}BND0: Undefined, \\ BND1-3: Reset\end{tabular} & \begin{tabular}[c]{@{}c@{}}BND0: Undefined, \\ BND1-3: Reset\end{tabular} \\ \hline
\end{tabular}
\caption{\simplex{} context behavior during repetitive initialization and finalization.}
\label{tab:repeat}
\end{table*}

\section{Implementation}
\label{sec:implementation}

\subsection{Components of \simplex}
Unfortunately, there is no means of directly accessing the MPX bounds registers via a \texttt{mov} instruction.
Each of GCC, ICC and Microsoft's Visual C++ compiler do offer intrinsics, although these are only available if a MPX runtime is available and providing bounds checking~\cite{Ramakesavan2016}.
This means it is not possible to use these intrinsics for accessing the bounds registers without also suffering the continual risk of a bounds check clobbering the bounds registers.
Therefore, within \simplex{}  we provide a system readiness check, a minimal runtime to enable and disable MPX execution without the additional overhead of bounds checking and a bounds lookup table, accessor and mutator functions, and a test suite to verify proper operation of the library.
We demonstrate usage of this functionality in Figure~\ref{fig:usage-listing}.

\begin{figure}[h!]
	\begin{lstlisting}[language=diff]
 #include <sys/stat.h>
+#include "simplex.h"

-static LBM_GridPtr srcGrid, dstGrid;

void MAIN_initialize( const MAIN_Param* param ) {
+	process_specific_init();

-	LBM_allocateGrid( (double**) &srcGrid );
-	LBM_allocateGrid( (double**) &dstGrid );
+	double* ptr;
+	LBM_allocateGrid(&ptr);
+	qsetbndl(BND0, (uint64_t) ptr);
+	ptr = 0;
+	LBM_allocateGrid(&ptr);
+	qsetbndl(BND1, (uint64_t) ptr);
+	ptr = 0;

-	LBM_initializeGrid( *srcGrid );
-	LBM_initializeGrid( *dstGrid );
+	LBM_initializeGrid( *((LBM_GridPtr)qgetbndl(BND0)) );
+	LBM_initializeGrid( *((LBM_GridPtr)qgetbndl(BND1)) );
}
void MAIN_finalize( const MAIN_Param* param ) {
-	LBM_freeGrid( (double**) &srcGrid );
-	LBM_freeGrid( (double**) &dstGrid );
+	double* p0 = (double*) qgetbndl(BND0);
+	double* p1 = (double*) qgetbndl(BND1);
+	LBM_freeGrid(&p0);
+	p0 = 0;
+	LBM_freeGrid(&p1);
+	p1 = 0;

+	process_specific_finish();
}
	\end{lstlisting}
\caption{A \texttt{diff} file example of modifications needed to store global pointers in bounds registers from the \texttt{lbm} benchmark. In this example, the global pointers {\tt srcGrid} and {\tt dstGrid} are placed in {\tt BND0} and {\tt BND1} respectively.}
\label{fig:usage-listing}
\end{figure}

\paragraph{System Readiness Check}
Although it is possible for a user to test whether their system can support MPX from the command line using commands such as \texttt{lscpu} and \texttt{sysctl}, a program must be able to verify readiness itself and abort further execution if it cannot prove its readiness.
This is because CPUs which do not support MPX will silently interpret these MPX instructions as NOPs.
We have isolated the necessary checks from the GCC 5.0 MPX runtime library.
These checks verify that \registerOne{CPUID}{14} is set (indicating that the CPU supports the MPX extension), and that \registerTwo{XCR0}{3}{4} are set (indicating that the CPU should include the MPX registers as part of a context save and restore).

\paragraph{Enabling and Disabling Functions}
We also provide a way of enabling and disabling MPX operations within both kernel mode and user mode applications.
This can be done by setting flags on the \register{BNDCFGS} and \register{BNDCFGU} registers  respectively.
\registerOne{BNDCFGx}{0} enables interpretation of the MPX instruction extension, and \registerOne{BNDCFGx}{1} enables bounds register preservation when legacy instructions are encountered.
Before these flags are set, we perform a system readiness check.
Unlike the GCC runtime, we do \textit{not} set \registerTwo{BNDCFGx}{63}{12} with the base address of the bounds table.
This minimizes startup overhead, and also provides a small measure of security since attempting to access the bounds table as a means of leaking the contents of a bounds register will result in a segmentation fault.

\paragraph{Accessor and Mutator Functions}
For each of the four bounds registers, a common accessor and mutator wrapper function provides a handle to the bounds register.
There are four varieties of each wrapper function: lower-half 64 bits only, upper-half 64 bits only, all 128 bits, and a ``quick'' lower-half only which does not attempt to save the upper-half nor clean the stack of any spilled values.
The applicable bounds register is selected through an enumerator with four values, thus corresponding ``{\tt BND0}'' to the integral value 0 and so forth.
Within each wrapper function is the necessary extended assembly statements to either set or get the values from the bounds register.
When writing to the bounds registers, the value to be written is marshaled from the function arguments into a sib-addressed \texttt{bndmk} instruction.
When getting, the bounds register is spilled onto the stack above the stack pointer without moving the stack pointer as there is no bounds register-to-general purpose register instruction.
This is accomplished using a \texttt{bndmov} instruction.
As mentioned above, all accessor functions except the quick variants will sanitize this value on the stack in case the value stored within the bounds registers is sensitive.
We have verified that our extended assembly statements to perform the sanitization are not optimized by either GCC or Clang through disassembly and manual inspection.
See Figure \ref{fig:mpx-context} for more information on data flows to and from the MPX context.

\subsection{Security Impact of the \simplex{} Implementation}
Canella et al. recently reported a variety of Meltdown transient execution attacks, one of which is the Meltdown-BR (Bounds Check Bypass) attack~\cite{Canella2018,Lipp2018}.
Dekel also describes a post-exploitation technique called BoundHook, which allows an attacker to cause a bounds check exception in a user-mode context, then catch the exception to gain control over the thread execution~\cite{Dekel2017}.
With both of these vulnerabilities, \simplex{} does not increase a program's attack surface because both require a \texttt{\#BR} exception to be raised in order to initiate exploitation.
Since \simplex{} does not use the \texttt{bndcl}, \texttt{bndcu}, or \texttt{bndcn} instructions, no such exception will be raised by our code.
Additionally, because BoundsHook requires that the attacker has also already compromised machine administrator rights, any attacker who can successfully execute a BoundsHook intrusion can simply observe and modify the MPX context without the need to further compromise \simplex.

Because \simplex{} can be used in multi-threaded applications, we must address the dangers that an attacker-controlled thread could victimize a thread using \simplex{} to interact with the MPX bounds registers.
There is a short time window where data being loaded from the bounds register to a system register is spilled onto the stack.
We provide one mitigation in that \simplex{} will zero out the memory used by the \texttt{bndmov} spill instruction immediately after copying to the destination register in all accessor functions except for \texttt{qgetbndl()} which is performance- rather than security-optimized.
Because this zeroing is not guaranteed to be atomic, there is still a small risk that the attacker-controlled thread with a pointer to the bottom of the victim thread's stack could read this memory in a race condition assisted by a scheduler interrupt sometime between the spill from the bounds register to the time the stack memory is sanitized.
We instrumented our library using a PAPI API~\cite{Browne2000} software defined event to measure the frequency of context switches within the \simplex{} accessor functions and discovered that such a sequence of events did not occur.
We hypothesize that this is because the accessor functions do not require any system calls and are very short-lived, and thus unlikely to trigger the scheduler's watchdog timer.

\section{Evaluation}
\label{sec:evaluation}

We conducted our evaluation on an 8-core Intel Core i7-7700K CPU at 4.20 GHz with 62.8 GiB RAM running Ubuntu 20.04 LTS and the Linux 5.4 kernel.
The system under evaluation conforms to POSIX.1-2017, and uses GNU libc and POSIX thread implementation version 2.27. 


\subsection{Benchmarks}
We authored three benchmark fixtures to evaluate whether \simplex{} attains performance that is comparable to using general purpose registers.


\subsubsection{Load-Store Benchmark} 
\begin{figure*}[!t]
\begin{tikzpicture}
\begin{axis}[
boxplot/draw direction=y,
ylabel={Rate (operations/second)},
ymode=log,
ymin=1e+5,ymax=1e+7,
height=6cm,
cycle list={{red},{blue}},
boxplot={
        draw position={1/3 + floor(\plotnumofactualtype/2) + 1/3*mod(\plotnumofactualtype,2)},
        box extend=0.3
},
x=2cm,
xtick={0,1,2,...,12},
x tick label as interval,
xticklabels={%
        {General\\{\scriptsize store load}},%
        {Segmentation\\{\scriptsize store load}},%
        {MMX\\{\scriptsize store load}},%
        {XMM\\{\scriptsize store load}},%
        {MPX\\{\scriptsize store load}},%
},
        x tick label style={
                text width=2.5cm,
                align=center
        },
]

\addplot+[
boxplot prepared={
  median=5.31e+6,
  upper quartile=5.91e+6,
  lower quartile=4.68e+6,
  upper whisker=5.91e+6,
  lower whisker=6.66e+5
},
] 
table [row sep=\\,y index=0] {
  data\\ 4.72e+6\\ 
};

\addplot+[
boxplot prepared={
  median=1.90e+6,
  upper quartile=2.23e+6,
  lower quartile=1.70e+6,
  upper whisker=5.91e+6,
  lower whisker=7.26e+5
},
] 
table [row sep=\\,y index=0] {
  data\\ 1.94e+6\\ 
};

\addplot+[
boxplot prepared={
  median=5.23e+6,
  upper quartile=5.74e+6,
  lower quartile=4.65e+6,
  upper whisker=5.88e+6,
  lower whisker=7.26e+5
},
] 
table [row sep=\\,y index=0] {
  data\\ 4.66e+6\\ 
};

\addplot+[
boxplot prepared={
  median=4.90e+6,
  upper quartile=5.91e+6,
  lower quartile=4.17e+6,
  upper whisker=5.91e+6,
  lower whisker=5.40e+5
},
] 
table [row sep=\\,y index=0] {
  data\\ 4.46e+6\\ 
};

\addplot+[
boxplot prepared={
  median=5.38e+6,
  upper quartile=5.91e+6,
  lower quartile=4.75e+6,
  upper whisker=5.91e+6,
  lower whisker=6.55e+5
},
] 
table [row sep=\\,y index=0] {
  data\\ 4.81e+6\\ 
};

\addplot+[
boxplot prepared={
  median=2.08e+6,
  upper quartile=2.43e+6,
  lower quartile=1.69e+6,
  upper whisker=4.86e+6,
  lower whisker=5.72e+5
},
] 
table [row sep=\\,y index=0] {
  data\\ 1.98e+6\\ 
};

\addplot+[
boxplot prepared={
  median=3.43e+5,
  upper quartile=3.48e+5,
  lower quartile=3.35e+5,
  upper whisker=3.52e+5,
  lower whisker=2.61e+5
},
] 
table [row sep=\\,y index=0] {
  data\\ 3.40e+5\\ 
};

\addplot+[
boxplot prepared={
  median=8.66e+5,
  upper quartile=8.91e+5,
  lower quartile=8.30e+5,
  upper whisker=1.04e+6,
  lower whisker=4.14e+5
},
] 
table [row sep=\\,y index=0] {
  data\\ 8.51e+5\\ 
};

\addplot+[
boxplot prepared={
  median=5.70e+6,
  upper quartile=5.91e+6,
  lower quartile=4.77e+6,
  upper whisker=5.91e+6,
  lower whisker=6.89e+5
},
] 
table [row sep=\\,y index=0] {
  data\\ 4.73e+6\\ 
};

\addplot+[
boxplot prepared={
  median=1.47e+6,
  upper quartile=1.69e+6,
  lower quartile=1.2e+6,
  upper whisker=2.55e+6,
  lower whisker=3.00e+5
},
] 
table [row sep=\\,y index=0] {
  data\\ 1.44e+6\\ 
};

\end{axis}
\end{tikzpicture}
\caption{Rate of load and store operations. Box and whisker plot shows median, minimum/maximum, and first/third quartile operation rates. We use \register{r15} for \textit{General}, \register{gs:0} for \textit{Segmentation}, \register{mm0} for \textit{MMX}, \register{xmm1} for \textit{XMM} and \register{bnd0} for \textit{MPX}. The test consisted of $10^4$ runs, with $10^6$ iterations per run. We report the steady-state rate of operations accomplished per second.}
\label{fig:load-store}
\end{figure*}
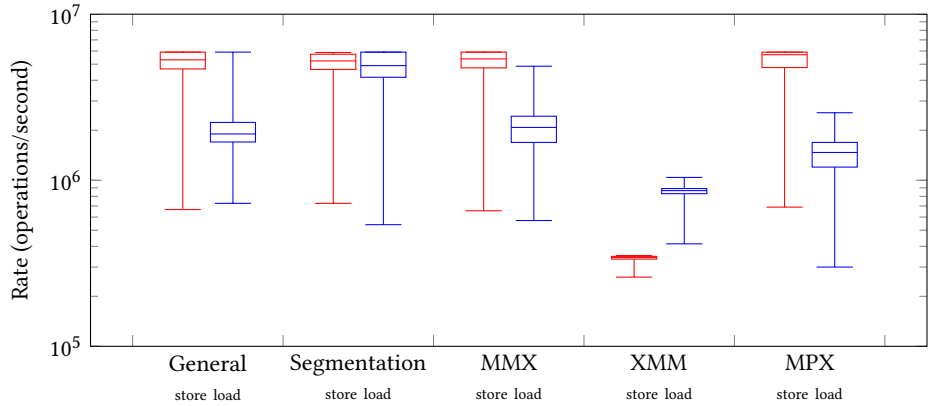

First, we authored a micro-benchmark that tests load and store performance when \simplex{} employs the \register{bnd0} MPX bounds register compared to handwritten assembly using general purpose registers using \register{r15}, segmentation registers using \register{gs:0}, and the MMX and XMM instruction set extension registers using \register{mm0} and \register{xmm1} respectively, see Figure~\ref{fig:load-store}.
We find that the mean of writing to the MPX bounds registers is comparable to writing to the general purpose registers ($1.00x$), segmentation registers ($1.01x$), and MMX registers ($0.98x$).
This is because all four of these operations have a fast, dedicated assembly instruction for writing to the register \-- either \texttt{mov} or \texttt{bndmk}.
The fastest assembly instruction option for writing to the XMM registers is \texttt{movaps}, which moves four aligned, packed, single-precision floating point values to the XMM register.
However, it incurs significant overhead compared to the \texttt{mov} instruction because of microarchitectural limitations and thus the rate of MPX bounds register writes is $13.90x$ faster.

Loading from the MPX bounds registers is a different story. 
Additional overhead results because the MPX extension does not contain an instruction to move from a bounds register directly to another register, whether a bounds register or otherwise; \texttt{bndmov} only provides a bounds register to memory spill operation.
Therefore load operations from a bounds register require that the data is first spilled to the quadword above the stack pointer through a \texttt{bndmov} instruction, then recovered through two additional \texttt{mov} instructions.
General purpose register, segmentation register and MMX register loads can all be accomplished by a single \texttt{mov} instruction and thus MPX bounds register loads are only $0.74x$, $0.32x$, and $0.73x$ as fast, respectively. Segmentation register loads are particularly fast when repeatedly executed because of cache effects. 
Conversely, MPX bounds register loads are $1.69x$ faster than XMM register loads because these loads also must spill to stack, and because of the aforementioned micro-architectural limitations of the \texttt{apsmov} instruction.

Our findings also confirm the micro-architectural analysis of Oleksenko et al.~\cite{Oleksenko2017} which found that it was not necessarily the MPX bounds operations that were particularly expensive, but the management of the bounds table through a two-level table lookup \--- particularly the \texttt{bndstx} and \texttt{bndldx} instructions.
\simplex{} uses neither of these instructions and thus avoids their overhead.

\subsubsection{Traversal Benchmark} 
We authored a benchmark simulating unhiding information from two hidden buffers by combining their contents, as first suggested by Shamir~\cite{Shamir1979}.
In our benchmark, a buffer is split into two halves, with two registers dedicated to pointing at the half-buffers.
These pointers are repeatedly indexed and de-referenced to traverse the buffer, which ranges in size from 4 KiB to 16 MiB (chosen to reflect common page sizes for target operating systems and the x86-64 ISA).
We repeat this unhiding traversal for 100 runs of 1000 iterations, and measure an elapsed time for all iterations to determine a geometric mean in order to calculate a performance overhead.
Because this requires two load operations from a register for each byte unhidden, performance overhead rapidly accumulates, in this case between 272.45\% and 282.17\% depending on buffer size.
See Table \ref{tab:traversal} for full results.

\begin{table}[!htb]
\centering
	\begin{tabular}{|l|r|r|r|r|r|r|}
		\hline
	 & \multicolumn{2}{r|}{{\bf General Purpose}} & \multicolumn{2}{c|}{{\bf \simplex{}}} & \multicolumn{2}{c|}{{\bf Overhead}} \\ \hline
		Size & \multicolumn{1}{c|}{ $\overline{X}$} & \multicolumn{1}{c|}{Med} & \multicolumn{1}{c|}{ $\overline{X}$} & \multicolumn{1}{c|}{Med} & \multicolumn{1}{c|}{ $\overline{X}$} & \multicolumn{1}{c|}{Med} \\ \hline
		4K & 1.1 & 1.0 & 3.9 & 3.9 & 272.5 & 278.1 \\ \hline
		8K & 2.1 & 2.1 & 7.8 & 7.8 & 279.0 & 279.7 \\ \hline
		1M & 266.5 & 265.9 & 1018.3 & 1018.1 & 282.2 & 282.9 \\ \hline
		16M & 4277.3 & 4274.5 & 16335.7 & 16334.1 & 281.9 & 282.1 \\ \hline
	\end{tabular}
	\caption{Performance overhead incurred during a simulation of traversing two buffers and combining their values to decode an information-hidden buffer (as in Shamir's secret hiding scheme). Each experiment is run on four buffer sizes for 1000 iterations and measured for elapsed time. Buffer size is expressed in bytes, measurements are expressed in seconds, and overhead in percentage. $\overline{X}= Mean, Med = Median$.}
	\label{tab:traversal}
\end{table}

\subsubsection{String Operations} 
Third, we implemented five memory operations from the \texttt{string.h} header, using reference implementations from the libgcc codebase.
We then reimplemented these functions for \simplex{} to replace any passed argument that contains the address of a buffer with calls to instead load the address from an applicable bounds register.
These benchmarks show that the performance cost of \simplex{} is easily amortized, as we found that the maximum overhead was only 5.86\%, and a 0.69\% overall geometric mean. 
In the specific case of these function implementations, benchmarks that do not short-circuit (i.e. memcpy, memmove and memset) are able to amortize the cost fully compared to functions that do short-circuit (i.e. memcmp, memchr).
We do not claim that there is a performance benefit to \simplex{}, simply that if there is a performance cost, it is small enough to be unnoticeable to the user and that it is offset by the utility of the additional registers provided by \simplex{}.

\begin{figure}[!h]
\begin{tikzpicture}[scale=0.9]
\begin{axis}[
symbolic x coords={memcmp,memcpy,memmove,memset,memchr},
major tick length=0cm,
xtick=data,
enlarge y limits={upper,value=0.2},
ymajorgrids,
yminorgrids,
ybar,
bar width = 6pt,
tick label style={/pgf/number format/fixed},
legend pos=outer north east,
ylabel=Overhead (\%)
]
\addplot coordinates {(memcmp,0.00) (memcpy,0.16) (memmove,0.57) (memset,1.20) (memchr,-0.01)};

\addplot coordinates {(memcmp,-0.02) (memcpy,0.14) (memmove,0.22) (memset,2.88) (memchr,0.00)};

\addplot coordinates {(memcmp,-0.08) (memcpy,-0.06) (memmove,0.22) (memset,1.44) (memchr,0.00)};

\addplot coordinates {(memcmp,2.29) (memcpy,-0.06)  (memset,1.46) (memchr,0.02)};
\legend{4 KiB,8 KiB,1 MiB,16 MiB}
\end{axis}
\end{tikzpicture}
\caption{String.h benchmarks' median performance overhead at varying buffer sizes.}
\label{fig:string}
\end{figure}
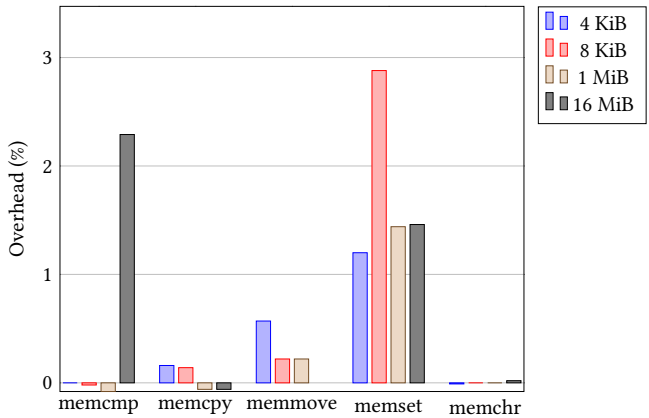

\begin{figure}[!h]
\centering
\scriptsize
\begin{tabular}{|l|r|r|r|r|r|r|r|r|r|r|}
	\hline
	\multicolumn{1}{|c|}{Size} & \multicolumn{2}{c|}{{\tt memcmp}} & \multicolumn{2}{c|}{{\tt memcpy}} & \multicolumn{2}{c|}{{\tt memmove}} & \multicolumn{2}{c|}{{\tt memset}} & \multicolumn{2}{c|}{{\tt memchr}} \\ \cline{2-11} 
	\multicolumn{1}{|c|}{} & \multicolumn{1}{c|}{$\overline{X}$} & \multicolumn{1}{c|}{Med} & \multicolumn{1}{c|}{$\overline{X}$} & \multicolumn{1}{c|}{Med} & \multicolumn{1}{c|}{$\overline{X}$} & \multicolumn{1}{c|}{Med} & \multicolumn{1}{c|}{$\overline{X}$} & \multicolumn{1}{c|}{Med} & \multicolumn{1}{c|}{$\overline{X}$} & \multicolumn{1}{c|}{Med} \\ \hline
	4K & 0.11 & 0.00 & 0.08 & 0.16 & 0.53 & 0.57 & 1.18 & 1.20 & 0.00 & -0.01 \\ \hline
	8K & -0.29 & -0.02 & 0.12 & 0.14 & 0.14 & 0.22 & 2.03 & 2.88 & -0.24 & 0.00 \\ \hline
	1M & 5.58 & -0.08 & -0.07 & -0.06 & -0.10 & -0.05 & 1.41 & 1.44 & 0.02 & 0.00 \\ \hline
	16M & 5.86 & 2.29 & -0.33 & -0.06 & - & - & 1.46 & 1.46 & 0.02 & 0.02 \\ \hline
\end{tabular}
\caption{String.h benchmarks' performance overhead when modified to pass pointer arguments in bounds registers.Performance overheads are expressed as percentages, while buffer sizes are expressed as kilobytes (K) or megabytes (M). $\overline{X}=Mean, Med = Median$}
\label{tab:string}
\end{figure}

\subsection{Modifications to Existing Codebases}
\subsubsection{SPEC CPU2017}
We hand-modified two SPEC \textsc{CPU2017} benchmarks, \texttt{519.lbm} which simulates fluid flow through lattices, and \texttt{531.deepsjeng} which plays chess.
In both cases, we selected the two global pointers to data structures that had the highest number of uses in order to fully stress the \simplex{} library.
Although we selected global objects, it should be emphasized that \simplex{} is not limited to just globals; heap or local objects could also be placed in the bounds registers.
Using the SPEC benchmarks proves both correctness \--- the output is verified against a known correct output \--- and demonstrates performance cost of using \simplex.
We measured the performance rate ratio between runs with an unmodified benchmark and one where frequently used pointers to global variables were placed into a bounds register.
This performance ratio was between 1.000 and 1.006 for \texttt{519.lbm}, and 0.975 and 0.985 for \texttt{531.lbm} (see Table \ref{tab:simplex spec}).
Higher performance rate ratios indicate faster execution, but differ from performance overhead measurements since performance rate takes into account the number of threaded copies running simultaneously. 

\begin{table*}[t]
\begin{tabular}{lrrrr}
\hline
Variables in Bounds Register & \multicolumn{1}{l}{Copies} & \multicolumn{1}{l}{\textbf{Run Time}} & \multicolumn{1}{l}{Base Rate} & \multicolumn{1}{l}{\textbf{Ratio}} \\ \hline
\multicolumn{5}{c}{519.lbm\_r}                                                        \\ \hline
None                                                         & 1 & 202 & 5.21 &       \\
None                                                         & 4 & 605 & 6.96 &       \\
srcGrid \textrightarrow{}bnd0                                & 1 & 201 & 5.24 & 1.006 \\
srcGrid \textrightarrow{}bnd0                                & 4 & 605 & 6.96 & 1.000 \\
srcGrid \textrightarrow{}bnd0, dstGrid \textrightarrow{}bnd1 & 1 & 202 & 5.23 & 1.004 \\
srcGrid \textrightarrow{}bnd0, dstGrid \textrightarrow{}bnd1 & 4 & 606 & 6.96 & 1.000 \\ \hline
\multicolumn{5}{c}{531.deepsjeng\_r}                                                  \\ \hline
None                                                         & 1 & 283 & 4.04 &       \\
None                                                         & 4 & 290 & 15.8 &       \\
state \textrightarrow{}bnd0, gamestate \textrightarrow{}bnd1 & 1 & 288 & 3.98 & 0.985 \\
state \textrightarrow{}bnd0, gamestate \textrightarrow{}bnd1 & 4 & 297 & 15.4 & 0.975 \\ \hline
\end{tabular}
\caption{\simplex{} SPEC CPU2017 evaluation data. \textit{Run time} refers to how long the benchmark took to complete. \textit{Base Rate} refers to the raw performance of this benchmark relative to the SPEC CPU2017 reference machine and thus provides insight into the underlying system under test. \textit{Ratio} refers to the ratio of the modified benchmark's performance to the unmodified benchmark's performance taking into account the number of copies running on a multi-threaded system. $ratio < 1$ implies the modified benchmark ran slower than the unmodified benchmark.}
\label{tab:simplex spec}
\end{table*}

\subsubsection{OpenSSL}
We then modified the OpenSSL Blowfish symmetric key cipher to demonstrate how \simplex{} might be used in a security application.
In our modified Blowfish cipher, the address of the cipher's global key schedule structure is stored in a bounds register.
Therefore wherever an encryption or decryption function would ordinarily receive a pointer to the key schedule as a function parameter, we instead pass a null value as the parameter and thus de-reference the bounds register at each usage of the parameter.
Although the OpenSSL test suite provides test run time in its output, the Blowfish correctness test is very short in duration.
As a result, our observed runtime overheads are smaller than the reported measurement resolution and not particularly useful as a metric of performance.
We do not wish to imply that replacing function parameters with \simplex{} function calls is a way to achieve higher performance, only to state that \simplex{} presents minimal performance cost.
We also wish to emphasize that although we placed a pointer to a key schedule structure in the bounds registers for this evaluation, this structure is stored on the heap in the unmodified Blowfish cipher and therefore we did not introduce attack surface in our modified cipher.
Additionally, some other OpenSSL ciphers' keys are less than 512 bits in size and would fit entirely within the bounds registers.
The MPX bounds registers can hold any value, not just pointer values.

\vspace{-.12in}
\subsubsection{DPlus}
Finally, we modified DPlus, a simple web browser written entirely in pure C++. 
The choice of implementation language is critical since \simplex{}  is not compatible with interpreted languages like Javascript, and the majority of modern browsers implement at least some portion of the code in Javascript for compatibility with popular web toolkits.
We verified correctness using the DW window toolkit's native functionality tests.


\section{Related Work}
\label{sec:related work}

\paragraph*{Existing Evaluations} Significant exploration of Intel MPX generally find MPX to be flawed as a memory safety tool, and thus inspired our investigation as to whether MPX could be repurposed.
Serebryany unfavorably evaluated the performance of Intel MPX versus the Address Sanitizer memory safety tool.~\cite{Serebryany2016}
Notably, he discovered not only up to a $2.5x$ performance slowdown and $4.0x$ memory overhead on some benchmarks, but that the MPX instructions still exhibit a 50\% slowdown even when they should be ignored on a system which does not have MPX support or has disabled it.
He also identifies three categories of false positives that Address Sanitizer does not have: atomic pointers, un-instrumented bounds changes, and those caused by compiler optimizations after instrumentation.
Otterstad examined the effectiveness of early implementations of MPX, identifying eight new categories of false positives and false negatives beyond those explored by Serebryany.~\cite{Otterstad2015}
Furthermore, he demonstrates at least one toy program which can be victimized by ROP attacks because of these false positives and false negatives. 
Oleksenko et al. performed a study of the performance, security guarantees, and usability issues of MPX after it became available in production hardware. ~\cite{Oleksenko2017}
Furthermore, their empirical study was backed by an exhaustive investigation of how MPX is actually implemented at the hardware, operating system and software levels.
This investigation is used to support their quantitative findings. 

\paragraph*{Other Uses of Intel MPX} We are not the only members of the community to propose repurposing MPX.
Code Pointer Integrity (CPI) maintains a safe region to protect function pointers, return addresses and other pointers to code called a ``safe stack''.~\cite{Kuznetsov2014}
The authors propose one implementation of CPI using MPX to store the safe region's metadata, gaining performance benefits by moving some of the implementation into MPX's hardware accelerated checks.
Burow further investigates using MPX to isolate CPI's shadow stacks and provide a highly-efficient implementation.~\cite{Burow2018}
We note that \simplex{} performs much of the management functionality they described, and could be used in conjunction with their defenses.
Opaque Control-Flow Integrity (O-CFI) combines fine-grained code layout randomization with coarse-grained CFI in order to defeat sophisticated attacks seeking in-memory layout information to launch code-reuse attacks.~\cite{Mohan2015}
O-CFI uses MPX instructions to perform branch instrumentation, where legal branch targets are ``chunked'' together into a minimal address range, similar to a buffer.
Oleksenko proposes a system combining MPX for hardware fault detection with Intel Transactional Synchronization Extensions (TSX) for fault rollback.~\cite{Oleksenko2016}
The underlying principle is that if a pointer's value is corrupted by a fault, then it will likely point to a dramatically different address outside the bounds of the referent object.
MemSentry is a deterministic memory isolation framework addressing the threats of allocation oracles, thread spraying, crash-resistant memory disclosure primitives, and various side channels.~\cite{Koning2017}
The authors use MPX and Intel Memory Protection Keys (MPK) to describe a more efficient method of intra-process isolation, similar to that provided by the kernel through \texttt{mprotect} and Software Fault Isolation (SFI).
CFIXX is a C++ defense for virtual table pointers providing Object Type Integrity (OTI).~\cite{Burow2018}
CFIXX protects against corruption attacks against OTI by protecting the memory region containing the OTI metadata with selective MPX instrumentation.
By reimagining the layout of the address space, they are able to halve the number of bound checks compared to a full memory safety solution provided by MPX.
BOGO extends the MPX bounds tables to not only provide spatial memory safety, but also temporal memory safety.~\cite{Zhang2019}
Since MPX already initializes bounds table entries at allocation, BOGO additionally invalidates these entries upon deallocation and thus gains temporal memory safety.
Since doing this operation at every deallocation can be expensive, the authors also introduce more efficient techniques for managing the deallocation metadata updates and for scanning the bounds table.
DataShield provides three methods for coarse-grained bounds checks for non-sensitive pointer dereferences, one of which utilizes MPX to avoid the need to information hide the non-sensitive data regions.~\cite{Carr2017}
Up to four of these regions' addresses are initialized in the MPX bounds register at program startup, with each pointer dereference in order to assure that the pointer does not escape the non-sensitive region.
The Linux kernel can be protected against Just-in-Time code reuse attacks by kR\^{}X, which hardens benign read operations that an attacker might reuse to disclose code to find useful JIT gadgets.~\cite{Pomonis2017}
Intel MPX is used in one implementation of kR\^{}X to accelerate the execute-only range checks to reduce the performance overhead.

\paragraph*{Repurposing Hardware Registers} The idea of repurposing hardware registers as with \simplex{} is not unique.
TRESOR is a patch that implements the AES encryption algorithm for the Linux kernel, but provides additional security by utilizing the Intel AES-NI instruction set extension and by keeping encryption keys instead of in RAM.~\cite{Muller2011}
Ginseng keeps secrets in an encrypted secure stack until they are needed, then moves the secret into dedicated registers.~\cite{Yun2019}
This has the effect of reducing the amount of sensitive data kept in the ARM TrustZone Trusted Execution Environment (TEE) and thus reduces the TEE's attack surface and does not require placing the operating system within the trusted computing base.


\section{Conclusion}
\label{sec:conclusion}

\simplex{} is an open-source library repurposing the Intel\textregistered MPX instruction set (ISA) extension.
We present evidence that suggests that MPX is ubiquitous, and show that MPX bounds registers can be repurposed as general purpose storage. In particular, they can be used to hide security sensitive data. 
We demonstrate that although the MPX ISA lacks a dedicated instruction to move data directly to and from the bounds registers, it is still possible to do so through the available spill and fill instructions, {\tt bndmk} and {\tt bndmov}.
Furthermore, we show that such operations are not overly-burdensome, especially once the operations are amortized across the entire execution of a program.
We do this through a collection of refactored programs and a partial implementation of the C standard library.

\section{Availability}
We make \simplex{} available to the community as open-source software at \simplexurl.

\bibliography{simplex}
\bibliographystyle{ACM-Reference-Format}

\end{document}